\documentclass[twocolumn,aps,prl,showpacs,amsmath]{revtex4-1}
\usepackage{amsfonts}
\usepackage{amssymb}
\usepackage{mathrsfs}
\usepackage{graphicx}
\usepackage{float}
\usepackage{xcolor}

\begin{document}

\title{Equivalence and superposition of real and imaginary quasiperiodicities}

\author{Xiaoming Cai}
\email{cxmpx@wipm.ac.cn}

\author{Shaojian Jiang}
\email{jiangsj@wipm.ac.cn}
\address{State Key Laboratory of Magnetic Resonance and Atomic and Molecular Physics, Wuhan Institute of Physics and Mathematics, APM, Chinese Academy of Sciences, Wuhan 430071, China}
\date{\today}

\begin{abstract}
We take non-Hermitian Aubry-Andr\'{e}-Harper models and quasiperiodic Kitaev chains as examples to demonstrate the equivalence and superposition of real and imaginary quasiperiodic potentials (QPs) on inducing localization of single-particle states. 
We prove this equivalence by analytically computing Lyapunov exponents (or inverse of localization lengths) for systems with purely real and purely imaginary QPs. 
Moreover, when superposed and with the same frequency, real and imaginary QPs are coherent on inducing the localization, under a way which is determined by the relative phase between them. 
The localization induced by a coherent superposition can be simulated by the Hermitian model with an effective strength of QP, implying that models are in the same universality class.
When their frequencies are different and relatively incommensurate, they are incoherent and their superposition leads to less correlation effects. 
Numerical results show that the localization happens earlier and there is an intermediate mixed phase lacking of mobility edge. 
%
\end{abstract}
\maketitle

\begin{figure}[tbp]
	\begin{center}
		\includegraphics[scale=0.4, bb=0 0 454 356]{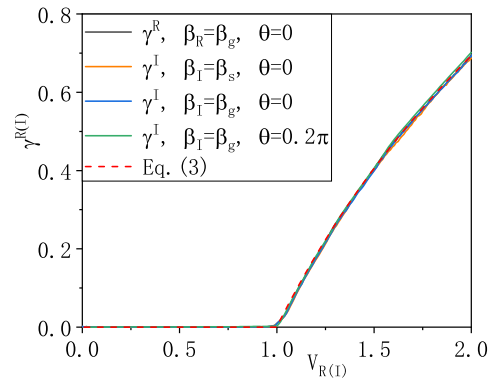}
		\caption{Equivalence in AAH models. 
			Mean LEs for systems with purely real and purely imaginary QPs. 
			Parameters: $\delta=0$, $\beta_g\simeq610/987$, $\beta_s\simeq408/985$, and $L=987$ or $985$.}
		\label{Fig1}
	\end{center}
\end{figure}

\begin{figure}[tbp]
	\begin{center}
		\includegraphics[width=\linewidth, bb=62 264 590 571]{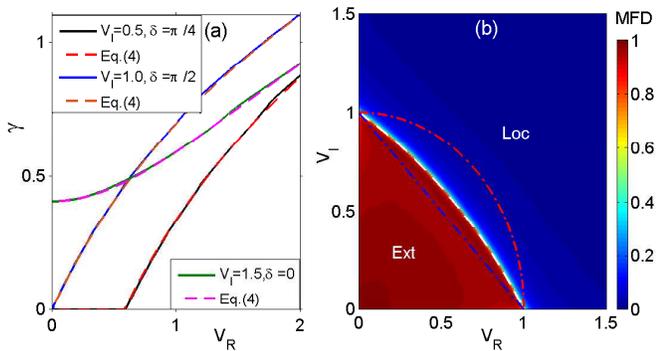}
		\caption{Superposition in AAH models: same frequency case. 
			(a) Mean LEs vs. $V_R$ for systems with different $V_I$ and $\delta$. 
			(b) Mean fractal dimension in the ($V_R,V_I$) plane for the system with $\delta=0.2\pi$. 
			White dash line represents Eq.(\ref{Condi}) under the condition $\delta=0.2\pi$, red dash dot line is described by equation $V_R^2+V_I^2=t^2$, and blue dash dot straight line corresponds to equation $V_R+V_I=t$. 
			Other parameters: $\beta_R=\beta_I=\beta_g\simeq610/987$, and $L=987$.}
		\label{Fig2}
	\end{center}
\end{figure}

Ever since the seminal work by P. W. Anderson in 1958, quantum localization has been a central topic in condensed matter physics \cite{Anderson1958,Abrahams2010}. 
According to scaling theory, an infinitesimal random disorder localizes all single-particle states in one dimension (1D) \cite{Evers2008}. 
However, as an intermediate case between disordered and periodic systems, quasiperiodic ones exhibit distinct behaviors and may support localization phase transitions. 
This is due to intrinsic spatial correlations of quasiperiodicities, which is absent for the true disorder.
A well-known example is the Aubry-Andr\'{e}-Harper (AAH) model \cite{Aubry1980}, which undergoes a transition from metal to insulator phases at a finite strength of the quasiperiodic potential (QP), guaranteed by a self-duality. 
The AAH model and its various extensions have been theoretically studied extensively \cite{Sarma1998,Cai2013,Rossignolo2019,Biddle2010,Ganeshan2015,Deng2019,Wang2020,Roy2021}, and experimentally realized in a variety of systems, such as ultracold atoms \cite{Roati2008,Li2017,Luschen2018,Kohlert2019,Goblot2020,An2021,An2018} and photonic crystals \cite{Lahini2009,Kraus2012}.
On the other hand, there has been growing interest in non-Hermitian physics recently \cite{Bender1998,Ashida2020}. 
Non-Hermiticity originates from exchanges of energy and/or particles with environment, and is embodied in the Hamiltonian as nonreciprocal hoppings, complex potentials, etc.
It leads to various unique phenomena, such as parity-time ($\mathcal{PT}$) symmetry breaking \cite{Mostafazadeh2010,Znojil2019}, non-Hermitian topology \cite{Gong2018,Song2019}, and skin effect \cite{Lee2016,Leykam2017,Shen2018,Yao2018,Kunst2018,Yokomizo2019,Borgnia2020,Okuma2020}.
Quantum localization has also been studied in non-Hermitian disordered \cite{Hatano1996,Silvestrov1999,Longhi2015,Freilikher1994,Asatryan1996,Basiri2014,Luo2021} and quasiperiodic \cite{Cai2020,Liu2020a,Liu2020b,Yuce2014,Liu2020c,Longhi2020a,Luo2019,Liang2014,Harter2016, Rivolta2017,Zeng2020a,Longhi2019a,Zhang2020b,Longhi2019b, Zeng2017,Jazaeri2001,Liu2020d,Liu2020e,Longhi2019} systems. 
By extending to the non-Hermitian realm, systems gain extra degrees of freedom and may host novel localization phenomena, such as nonreciprocal hopping induced delocalization \cite{Hatano1996,Silvestrov1999,Longhi2015}, and new universality classes of Anderson transitions \cite{Luo2021}. 
However, the relation and difference between localizations in Hermitian and non-Hermitian systems are unrevealed yet.
In particular, the extension of QP from real to complex results in various AAH models with both real and imaginary QPs. 
Although several special cases have been studied before \cite{Cai2020,Longhi2019,Longhi2019b,Liu2020f}, showing similar localization behaviors, the extension naturally raises the following fundamental questions. 
First, what are the roles on inducing localization of states, played by purely real and corresponding imaginary QPs?
Are they simply equivalent? 
Note that they have distinct physical origins, and spectra for systems with them can be quite different.
Second, when superposed, whether real and imaginary parts of complex QPs are coherent on inducing the localization? 
How does the superposition affect the spatial correlations?
%
%
Furthermore, is it possible to simulate in the Hermitian system, the localization induced by a superposition?
In the paper, we attempt to address above questions by studying the localization in 1D non-Hermitian AAH models with both real and imaginary QPs. 
Applying Avila's global theory of one-frequency Schr\"{o}dinger operators \cite{Avila2015} and Thouless's result \cite{Thouless1}, we analytically prove that purely real and imaginary QPs result in the same localization length and phase transition point, which implies that they are equivalent on inducing localization of states.  
%
%
%
When superposed, superposition principles for real and imaginary QPs with the same frequency are analytically established. 
Meanwhile, the localization induced by their coherent superposition is examined. 
Furthermore, the incoherent superposition of them with different and relatively incommensurate frequencies is numerically studied, which shows less correlation effects.
\emph{Model.--} We consider the 1D non-Hermitian AAH model with and without $p$-wave pairing, described by the following Hamiltonian
\begin{equation}
H=\sum_j(-tc^\dagger_jc_{j+1}+\Delta c_jc_{j+1}+\mathrm{h.c.})+\sum_jV_jc^\dagger_jc_j,
\label{Ham}
\end{equation}
where $c^\dagger_j(c_j)$ is the creation(annihilation) operator of a spinless fermion at site $j$. 
$t$ is the hopping amplitude and sets the unit of energy ($t=1$). 
$\Delta$ is the $p$-wave pairing amplitude, which can be made positive real \cite{Kitaev2001}.
We choose a general form of QP, which is given by
\begin{equation}
V_j=2V_R\mathrm{cos}(2\pi\beta_R j+\theta)+2iV_I\mathrm{cos}(2\pi\beta_I j+\theta+\delta),
\end{equation}
with $V_R$ and $V_I$ strengths of the real and imaginary QPs, respectively. 
$\theta$ is a global phase, which is trivial on the localization, and we will set $\theta=0$ if not specified. 
$\delta$ is the relative phase between the real and imaginary QPs. 
In the presence of phases $\theta$ and $\delta$, we can set both $V_{R}$ and $V_{I}$ positive real. 
$\beta_{R}$ and $\beta_{I}$ are irrational numbers characterizing quasiperiodicities of the real and imaginary QPs, respectively. 
In this paper, we choose the metallic mean family \cite{Falcon2014} of irrational numbers. 
Considering a generalized $k$-Fibonacci sequence given by $F_m=kF_{m-1}+F_{m-2}$ with $F_0=0$ and $F_1=1$, the limit $\beta=\lim_{m\rightarrow\infty}F_{m-1}/F_m$ with $k=1,2,3...$ yields the metallic mean family. 
The first three members of the family, which are used in this paper, are the well-known golden mean $\beta_g=(\sqrt{5}-1)/2$, the silver mean $\beta_s=\sqrt{2}-1$, and the bronze mean $\beta_b=(\sqrt{13}-3)/2$, respectively. 
Each number of the family satisfies the relation $k\beta+\beta^2=1$.
The localization phenomena on few specific lines in ($V_R,V_I$) plane have been studied before, for models with $\beta_R=\beta_I=\beta$ and $\delta=-\pi/2$. 
For example, the potential $2V\mathrm{cos}(2\pi\beta j+ih)$ studied in Ref.\cite{Cai2020,Longhi2019,Cai2021} corresponds to $V_R=V\mathrm{cosh}(h)$ and $V_I=V\mathrm{sinh}(h)$; the potential $2Ve^{2\pi\beta j}$ in Ref.\cite{Longhi2019b,Liu2020f} has $V_{R(I)}=V$; and the potential $V_R\mathrm{cos}(2\pi\beta j)+iV_I\mathrm{sin}(2\pi\beta j)$ was numerically studied in Ref.\cite{Jazaeri2001}. 
However, the study is still missing, on general localization and more importantly the fundamental principles behind, which are focuses of the paper.
\emph{Equivalence and superposition in AAH models.--} We first examine the case without $p$-wave pairing ($\Delta=0$), which is a typical non-Hermitian AAH model. 
To prove the equivalence between purely real and imaginary QPs on inducing the localization, we analytically compute Lyapunov exponents (LEs) (or inverse of localization lengths) of single-particle states. 
Given an eigenstate $|\Phi\rangle=\sum_j\phi_jc^\dagger_j|0\rangle$, the Schr\"{o}dinger equation of amplitudes $\phi_j$ in transfer matrix form is written as
\begin{equation}
\left[\begin{array}{c}
\phi_{j+1}\\
\phi_j\end{array}
\right]=T_j\left[\begin{array}{c}
\phi_{j}\\
\phi_{j-1}\end{array}
\right],T_j=\left[\begin{array}{cc}
(V_j-E)/t&-1\\
1&0\end{array}
\right],\notag
\end{equation}
with $E$ the eigenenergy. 
The LE of state is computed by
\begin{equation}
\gamma_\varepsilon=\lim_{L\rightarrow\infty}\frac{1}{L}\mathrm{ln}\|\prod_{j=1}^LT_j(\theta+i\varepsilon)\|,\notag
\end{equation}
where $\|\cdot\|$ denotes the norm of a matrix, and $L$ is the number of lattice sites.
Since our computation relies on Avila's global theory of one-frequency analytical $SL(2,\mathbb{C})$ cocycle \cite{Avila2015}, complexification of the global phase $\theta$ in $T_j$ has been performed. 
Let $\varepsilon$ go to infinity, then a direct calculation yields transfer matrices
\begin{equation}
T_j^{R(I)}(\varepsilon)=e^{\varepsilon-i2\pi\beta_{R(I)}j -i\theta}\left[\begin{array}{cc}
V_R(iV_Ie^{-i\delta})/t&0\\
0&0\end{array}
\right]+o(1),\notag
\end{equation} 
for purely real and imaginary QPs, respectively. 
Thus we obtain $\gamma^{R(I)}_{\varepsilon\rightarrow\infty}=\varepsilon+\ln|V_{R(I)}/t|$. 
According to Avila's global theory, as a function of $\varepsilon$, $\gamma_\varepsilon$ is a convex, piecewise linear function with integer slopes.
Moreover, the energy $E$ does not belong to the spectrum, if and only if $\gamma_{\varepsilon=0}(E)>0$ and $\gamma_\varepsilon$ is an affine function in a neighborhood of $\varepsilon=0$. 
It follows that
\begin{equation}
\gamma^{R(I)}=\max\{\ln|V_{R(I)}/t|,0\},
\label{Equv}
\end{equation}
for AAH models with purely real and purely imaginary QPs, respectively. 
The LEs do not depend on the energy $E$, frequencies $\beta_{R/I}$, and phases $\theta$ and $\delta$. 
More importantly, purely real and imaginary QPs with the same strength have exactly the same LE, implying that they are equivalent on inducing localization of states. 
When $V_{R(I)}/t<1$, the LE $\gamma^{R(I)}=0$ and states are extended, whereas states are localized when $V_{R(I)}/t>1$. 
The localization phase transition happens at $V_{R(I)}/t=1$. 
The equivalence and parameter-(in)dependence are further verified numerically. 
We adopt exponentially decaying wave functions $\phi^n_j=\mathrm{exp}(-\gamma_n|j-j_0|)$ with $j_0$ the localization center, $n$ the index of states, and $\gamma_n$ the LE.
Extracted by fitting numerical single-particle states with the above wave functions, the mean LEs $\gamma=\sum_n\gamma_n/L$ for systems with purely real and purely imaginary QPs are shown in Fig.\ref{Fig1}. 
All curves coincide with the theoretical prediction in Eq.(\ref{Equv}).
Having established the equivalence, we study the superposition of real and imaginary QPs with the same frequency ($\beta_R=\beta_I=\beta$).
By applying Avila's global theory, we obtain the LE in the presence of both QPs
\begin{equation}
\gamma=\max\{\frac{1}{2}\ln\frac{V^2_R+V^2_I+2V_RV_I\arrowvert\mathrm{sin}\delta\arrowvert}{t^2},0\},
\label{Supp}
\end{equation}
[see Supplemental Material (SM) for the analytical derivation] \cite{SM}. 
Thus, localization phase transition points are determined by condition
\begin{equation}
V^2_R+V^2_I+2V_RV_I\arrowvert\mathrm{sin}\delta\arrowvert=t^2.
\label{Condi}
\end{equation}
These clearly reflect the coherent superposition of real and imaginary QPs, where their relative phase $\delta$ plays a decisive role. 
The LE and condition for phase transition are periodic functions of the phase $\delta$ with a period $\pi$. 
Moreover, they are symmetric with respect to $\delta=m\pi/2, m\in\mathbb{Z}$ . 
When $\delta=0$, we have LE $\gamma=\ln\sqrt{V_R^2+V_I^2}/t$ in the localized phase, and phase transition points are determined by $V_R^2+V_I^2=t^2$, which is part of the unit circle [red dash dot line in Fig.\ref{Fig2}(b)]. 
When $\delta=\pi/2$, the LE $\gamma=\max\{\ln(V_R+V_I)/t,0\}$, and phase transitions occur at $V_R+V_I=t$ [blue dash dot straight line in Fig.\ref{Fig2}(b)]. 
With a general $\delta$, the phase transition line varies between the above two cases back and forth. 
The coherent superposition induces the localization transition earlier [at smaller magnitudes of the vector ($V_R,V_I$)] than purely real or imaginary QP solely, implying that the superposition leads to less correlation effects.
Effects of real and imaginary QPs are equal in the superposition.
Additionally, near a phase transition point the LE in localized phase scales linearly with the distance from the phase transition point on the $(V_R,V_I)$ plane. 
Thus the non-Hermitian AAH model considered here is in the same universality class as the Hermitian one, with a localization exponent $1$ \cite{Aubry1980,Sarma1998}. 
Furthermore, note that Eqs.(\ref{Equv}) and (\ref{Supp}) are in the same form, and the localization induced by a superposition of real and imaginary QPs can be simulated by the Hermitian AAH model with an effective strength of QP [$V_{\mathrm{eff}}=\sqrt{V^2_R+V^2_I+2V_RV_I\arrowvert\mathrm{sin}\delta\arrowvert}$].

The above analytical results are consistent with numerical simulations. 
In Fig.\ref{Fig2}(a) we show examples of mean LEs vs. $V_R$ for systems with different $V_I$ and $\delta$, which agree with Eq.(\ref{Supp}). 
To explore the localization in more details, we further compute inverse participation ratios (IPRs) and fractal dimensions of single-particle states. 
For a normalized state the IPR is defined as $P=\sum_j|\phi_j|^4$. 
In general, the IPR $P\propto L^{-\alpha}$ with a fractal dimension $0\leq\alpha\leq1$. 
For an extended state, $P\propto 1/L$ and $\alpha=1$, whereas the IPR approaches $1$ and $\alpha=0$ for a localized state. 
States with $0<\alpha<1$ are critical, and have multi-fractal properties. 
Extracted by the box-counting method \cite{SM}, a typical mean fractal dimension $\mathrm{MFD}=\sum_n\alpha_n/L$ in $(V_R,V_I)$ plane is shown in Fig.\ref{Fig2}(b). 
When both $V_R$ and $V_I$ are small, the system is in the extended phase with $\mathrm{MFD}\simeq1$, whereas it is in the localized phase with $\mathrm{MFD}\simeq0$ when $V_R$ and/or $V_I$ are large. 
Critical states only exist at the boundary between two phases, which agrees with the theoretical prediction in Eq.(\ref{Condi}) [white dash line in Fig.\ref{Fig2}(b)].
\begin{figure}[tbp]
	\begin{center}
		\includegraphics[width=\linewidth, bb=8 262 542 566]{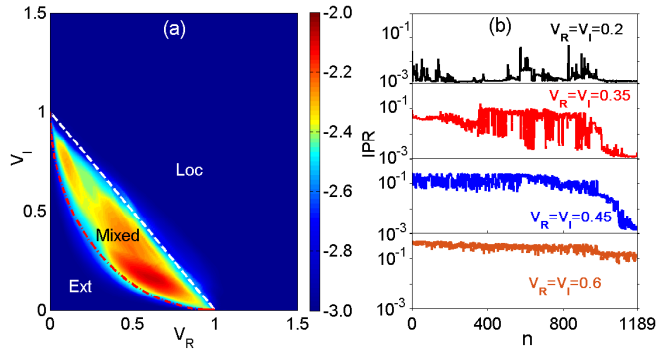}
		\caption{Superposition in AAH models: two-frequency case. 
			(a) The quantity $\kappa$ [Eq.(\ref{MINPR})] in ($V_R,V_I$) plane. 
			White dash line corresponds to equation $V_R+V_I=t$, and the red dash dot one corresponds to the lower left quarter of  circle $(V_R-t)^2+(V_I-t)^2=t^2$. 
			(b) Distributions of IPRs for systems with different ($V_R,V_I$). 
			Parameters: $\beta_R=\beta_g\simeq610/987$, $\beta_I=\beta_b\simeq360/1189$, $\delta=0$, and $L=1189$.}
		\label{Fig3}
	\end{center}
\end{figure}
The superposition of real and imaginary QPs with different frequencies only can be numerically studied. 
We focus on the case where two frequencies are incommensurate to each other, chosen from the metallic mean family. 
The localization is basically independent of frequencies $\beta_{R/I}$ and phases $\theta$ and $\delta$ \cite{SM}, showing the incoherent nature of the superposition. 
Quantities characterizing the localization, such as the mean IPR, MFD, and mean LE, are functions of ($V_R,V_I$) only, and a typical one is shown in Fig.\ref{Fig3}(a), which is defined as \cite{Roy2021}
\begin{equation}
\kappa=\log_{10}(\mathrm{MIPR*MNPR}),
\label{MINPR}
\end{equation}
where $\mathrm{MIPR}=\sum_nP_n/L$ is the mean IPR.
The normalized participation ratio is defined as the `inverse' of IPR for a single-particle state, and the mean one is $\mathrm{MNPR}=\sum_n1/(P_nL^2)$. 
The quantity $\kappa$ is introduced to distinguish the mixed phase (the intermediate region bounded by dash lines, where $\kappa$ is finite) from the fully extended and localized phases (blue regions, where $\kappa\propto 1/L$). 
In the mixed phase, the system consists of extended, localized, and even critical single-particle states. 
The boundary between mixed and localized phases is well described by the empirical equation $V_R+V_I=t$ [white dash straight line in Fig.\ref{Fig3}(a)], while the lower left quarter of circle $(V_R-t)^2+(V_I-t)^2=t^2$ [red dash dot line in Fig.\ref{Fig3}(a)] approximately describes the boundary between extended and mixed phases. 
Compared with the same frequency case [Fig.\ref{Fig2}(b)], the superposition of real and imaginary QPs with different frequencies induces the localization earlier, thus has weaker correlation effects. 
To further explore details of the localization and the possible existence of mobility edges, we present distributions of IPRs in Fig.\ref{Fig3}(b) for systems with different strengths of QPs. 
States are arranged in ascending order of the real part of energies. 
When both $V_{R}$ and $V_I$ are small, the system is in the extended phase with $P\propto 1/L$. 
However, numerically there are few localized defect states caused by the mismatch of rational approximations of $\beta_{R}$ and $\beta_{I}$, and cusps exist in the upper panel of Fig.\ref{Fig3}(b). 
In the mixed phase, the IPR varies from orders of $1$ to $1/L$, which indicates that there is no mobility edge and the system consists of all three kinds of states.
In addition, a real superposition of QPs with relatively incommensurate frequencies [$V'_j=2V_1\mathrm{cos}(2\pi\beta_1 j+\theta)+2V_2\mathrm{cos}(2\pi\beta_2 j+\theta+\delta)$] leads to very similar localization behaviors as for the superposition studied here, implying that there likely is a correspondence between two models \cite{SM}.

\begin{figure}[tbp]
	\begin{center}
		\includegraphics[width=\linewidth, bb=22 263 585 583]{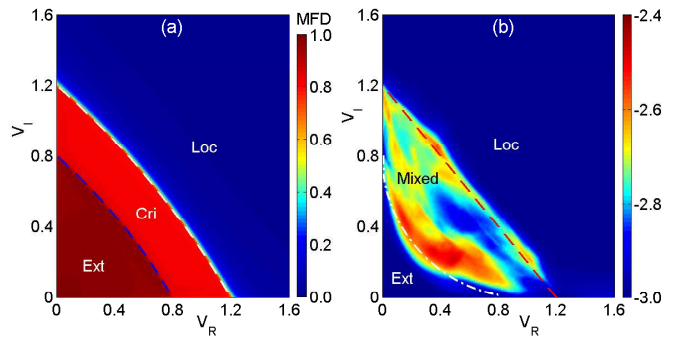}
		\caption{Superposition in quasiperiodic Kitaev chains. 
			(a) Mean fractal dimension in the $(V_R,V_I)$ plane for a system with the same frequency. 
			Parameters: $\Delta=0.2$, $\delta=\pi/4$, $\beta_R=\beta_I=\beta_g\simeq610/987$, and $L=987$. 
			White and blue dash lines correspond to Eqs.(\ref{CondiK1}) and (\ref{CondiK2}), respectively. 
			(b) Quantity $\kappa$ for the system with two frequencies. 
			Two lines are added, corresponding to equations $V_R+V_I=t+\Delta$ (red dash line) and $(V_R/|t-\Delta|-1)^2+(V_I/|t-\Delta|-1)^2=1$ (white dash dot line). 
			Parameters: $\Delta=0.2$, $\delta=0$, $\beta_R=\beta_g\simeq610/987$, $\beta_I=\beta_b\simeq360/1189$, and $L=1189$.}
		\label{Fig4}
	\end{center}
\end{figure}

\emph{Equivalence and superposition in quasiperiodic Kitaev chains.--} 
Turning on the $p$-wave pairing ($\Delta>0$), the model can be thought of as the Kitaev chain subjected to both real and imaginary QPs. 
It can be diagonalized by the BdG transformation $\eta^\dagger=\sum_j[\phi_j\chi^A_j+i\psi_j\chi^B_j]$, where $\chi^{A(B)}$ are Majorana operators, and $\phi$ ($\psi$) play the role of single-particle states \cite{SM}. 
By generalizing to the non-Hermitian realm Thouless's result relating LE to the density of state \cite{Thouless1}, the LEs can be analytically computed as
\begin{equation}
\gamma_K^{R(I)}=\max\{\ln\frac{V_{R(I)}}{t+\Delta},0\},
\label{EquvK}
\end{equation}
for quasiperiodic Kitaev chains with purely real and purely imaginary QPs, respectively \cite{SM}. 
Purely real and imaginary QPs are still equivalent on inducing the localization, and a numerical verification is presented in SM \cite{SM}. 
The LEs are independent of the energy $E$, frequencies $\beta_{R/I}$ and phases $\theta$ and $\delta$. 
When $V_{R(I)}>t+\Delta$, the system is in the localized phase with $\gamma_K>0$. 
However, it is not necessarily in the extended phase when $V_{R(I)}<t+\Delta$, despite $\gamma_K=0$. 
There is an extra transition point separating extended and critical phases at $V_{R(I)}=|t-\Delta|$. 
In the intermediate critical phase, states have fractal dimensions $0<\alpha<1$ [see the vertical and horizontal axes of Fig.\ref{Fig4}(a)]. 
%
%

%
When real and imaginary QPs are superposed and have the same frequency, the non-Hermitian extension of Thouless's result leads to the LE for quasiperiodic Kitaev chain \cite{SM}
\begin{equation}
\gamma_K=\max\{\ln\frac{\sqrt{V^2_R+V^2_I+2V_RV_I\arrowvert\mathrm{sin}\delta\arrowvert}}{t+\Delta},0\}.
\label{SuppK}
\end{equation}
Localization phase transition points are determined by condition
\begin{equation}
V^2_R+V^2_I+2V_RV_I\arrowvert\mathrm{sin}\delta\arrowvert=(t+\Delta)^2.
\label{CondiK1}
\end{equation}
Comparing Eqs.(\ref{Supp}) and (\ref{Condi}) with above two equations, respectively, the difference is that $t$ is replaced by $t+\Delta$.
Thus conclusions drawn in AAH models still hold for the quasiperiodic Kitaev chain. 
The LE and condition for phase transition are verified by numerical simulations, which are presented in SM \cite{SM}, along with other localization properties. 
Here, an example of mean fractal dimension is shown in Fig.\ref{Fig4}(a). 
Different from AAH models, there is a critical phase between extended and localized phases. 
The location of extended-critical phase transition is well described by equation
\begin{equation}
V^2_R+V^2_I+2V_RV_I\arrowvert\mathrm{sin}\delta\arrowvert=(t-\Delta)^2,
\label{CondiK2}
\end{equation}
which is similar to Eq.(\ref{Condi}) or Eq.(\ref{CondiK1}).

Finally, we study the superposition of QPs with two relatively incommensurate frequencies. 
The localization is still independent of $\theta$, $\delta$, and $\beta_{R/I}$, and thus QPs are incoherent. 
In Fig.\ref{Fig4}(b) we present the quantity $\kappa$ in $(V_R,V_I)$ plane. 
Two lines are added, corresponding to empirical equations that are similar to the ones for AAH models and obtained by substituting $t$ with $t\pm\Delta$.
The intermediate region corresponds to the mixed phase, and the localization happens before the extended-critical phase transition shown in the same frequency case. 
The superposition of real and imaginary QPs with different frequencies leads to weaker correlations. 
As strengths of QPs increase, more and more states become localized in the mixed phase \cite{SM}.
\emph{Conclusion and discussion.--} 
Through our work proves the equivalence between purely real and imaginary QPs, and studies the superposition of them on inducing localization of states. 
Specifically, LEs are analytically computed for AAH models by applying Avila's global theory, and for quasiperiodic Kitaev chains by generalizing to non-Hermitian realm Thouless's result relating LE to the density of state. 
Purely real and imaginary QPs induce the same localization. 
In the presence of both and with the same frequency, they are coherent on inducing the localization, under a way which is determined by the relative phase between them. 
While with different and relatively incommensurate frequencies, they  are incoherent and less correlated, and induce the localization earlier. 

The physics of equivalence and superposition can be experimentally tested in photonic waveguides and electric circuits. 
Photonic waveguides have been routinely used to demonstrate the localization of light \cite{Schwartz2007,Lahini2008}. 
In the tight-binding limit, propagation of classical light is governed by $i\mathrm{d}\phi_j/\mathrm{d}z=\kappa_j\phi_j+\sum_{l\neq j}t_{j,l}\phi_l$, which resembles the Schr\"{o}dinger equation. 
$\kappa_j$ is the refractive index of $j$th waveguide, which plays the role of complex potential. 
$t_{j,l}$ is the hopping between different waveguides. 
In photonics, the equation specified for the single-particle physics of Kitaev chain is even termed `photonic Kitaev chain' \cite{Xu2016_1}. 
In electric circuits the single-particle eigenvalue problem is simulated by the Kirchhoff's current law $I_a=\sum_{b=1}^LJ_{ab}V_b$, where the Laplacian of circuit $J$ acts as the effective Hamiltonian, and $I_a$ and $V_a$ are the current and voltage at node $a$ \cite{Ashida2020}, respectively. 
On-site complex potentials are realized by grounding nodes with proper resistors. 
It would be interesting to extend the study of equivalence and superposition to other systems, such as in the presence of disorders, mobility edges, and even interactions, where non-Hermitian many-body localization occurs.
\emph{Acknowledgments.--} This work is supported by the National Key R\&D Program of China under Grant No. 2017YFA0304500 and No. 2016YFA0301503, the NSFC grants No. 11534014 and No. 11874393.

\end{document}